\documentclass[apj, numberedappendix]{emulateapj}
\usepackage{epsfig,amsmath} 
\usepackage{graphicx}
\usepackage{txfonts}
\usepackage{amssymb}
 
\slugcomment{Received 1997 October 13; accepted 1998 August 3}
\journalinfo{The Astrophysical Journal, 510: 312:324, 1999 January 1}

\def\cminvsq{\,{\rm cm}^{-2}}

\def\s{{\rm s}}
\def\ms{\,{\rm ms}}

\def\ph{\,{\rm ph}}
\def\secinv{\,{\rm s}^{-1}}

\def\trd{t_{r,d}}
\def\taurd{\tau_{r,d}}
\def\tr{t_r}
\def\td{t_d}
\def\taur{\tau_r}
\def\taud{\tau_d}
\def\t0{t_0}

\newcommand{\bez}{\begin{eqnarray*}}
\newcommand{\eez}{\end{eqnarray*}}
\newcommand{\be}{\begin{equation}}
\newcommand{\ee}{\end{equation}}
\newcommand{\beq}{\begin{eqnarray}}
\newcommand{\eeq}{\end{eqnarray}}
\newcommand{\bc}{\begin{center}}
\newcommand{\ec}{\end{center}}

\begin{document}

\title{A Complexity-Brightness Correlation in Gamma Ray Bursts} 

\shorttitle{Turn-over in the Brightness Distribution of Gamma-Ray Bursts}
\shortauthors{STERN, POUTANEN, \& SVENSSON}



\author{Boris Stern \altaffilmark{1,2}, Juri Poutanen\altaffilmark{2}, 
and Roland Svensson\altaffilmark{2}} 

\altaffiltext{1}{Institute for Nuclear Research, Russian Academy of Sciences
Moscow 117312, Russia } 
\altaffiltext{2}{Stockholm Observatory, SE-133 36 Saltsj\"obaden, Sweden}


\begin{abstract}
We observe strong correlations between the temporal properties of gamma ray bursts 
(GRBs) and their
apparent peak brightness. The strongest effect (with a significance level of
$\sim 10^{-6}$) is  the difference 
between the brightness distributions  of simple bursts (dominated by a single 
smooth pulse) and complex bursts (consisting of overlapping pulses).
The latter has a break at a peak flux of $\sim 1.5 \ph \cminvsq \secinv$, while 
the distribution of simple bursts is smooth down to the BATSE threshold. 
We also observe brightness dependent variations in the 
shape of the average peak aligned time profile (ATP) of GRBs. The decaying slope
of the ATP shows time dilation when comparing bright and dim bursts while the 
rising slope hardly changes. Both slopes of the ATP are deformed for weak bursts 
as compared to strong bursts. The interpretation of these effects is simple:
a complex burst where a number of independent pulses overlap in    
time appears intrinsically stronger than a simple burst. Then the BATSE 
sample of complex bursts covers larger redshifts where some cosmological 
factor causes the break in the peak brightness distribution. 
This break could correspond to the peak in the star formation rate that was recently
shown to occur at a redshift of $z \sim 1.5$.
\end{abstract}

\keywords{gamma rays: bursts -- gamma rays: observations -- 
methods: statistical}

\section{Introduction}

Do the temporal properties of GRBs have systematic trends that are 
dependent on brightness? The most intensively discussed 
and studied trend is the time dilation of weak bursts relative to strong bursts.
The attention to the time dilation effects is due to its possible 
cosmological interpretation (Paczy\'nski 1992). 
The results of time dilation measurements
differ for different groups: Norris et al. (1994), 
Fenimore \& Bloom (1995), and
Stern (1996) found time dilation, while Mitrofanov et al. (1996) 
and Lee \& Petrosian (1997) did not find any significant effect. 

Positive detections of time dilation are consistent with the simplest 
assumption that weak events are just redshifted (and therefore stretched) 
analogs of strong events. However, in the work of Stern, Poutanen \& Svensson 
(1997, hereafter SPS97), it was demonstrated that the situation is more 
complicated. Weak events are not only longer on average, 
but they are also more asymmetric on average. Furthermore, their average 
peak aligned time profile (ATP) has a different shape as compared to 
strong events. This difference can be quantified
in terms of different stretched exponential indices.

Such brightness dependent correlations are in contradiction with a 
straightforward 
interpretation of the temporal stretching as being due to 
cosmological time dilation. 
On the other hand, these correlations can easily be interpreted
in terms of correlations between temporal properties and {\it intrinsic 
peak luminosity}. Indeed, GRBs are composed of individual asymmetric pulses with
fast rise and slower exponential decay (FREDs). 
The ATP of simple GRBs consisting of one pulse (or a few pulses) is on 
average more 
asymmetric than the ATP for complex GRBs where a chaotic 
bunch of overlapping pulses can produce an arbitrary time profile and 
the asymmetry related to the individual pulses is washed out. 
If we assume that different elementary pulses originate in different regions
being associated with local events in the course of a global event, 
then the amplitudes of elementary pulses with overlapping arrival time sum up. 
Therefore, complex bursts consisting of many overlapping pulses are intrinsically 
brighter than simple events. 
A direct morphological classification performed by SPS97 confirmed this interpretation:
complex bursts show systematically larger peak fluxes than simple GRBs.

Although such correlations are natural in intrinsic peak luminosity, we, however, 
observe them in a narrow range (within one decade) of {\it apparent} peak 
brightness (i.e., peak count rate). 
This means that the distribution
of GRBs over luminosity distance differs significantly from a power law. 
(In the case of a power law, intrinsically strong and intrinsically weak GRBs 
would be blended in the same proportion at any apparent brightness.)
These correlations can put additional constraints on the spatial 
distribution of the GRB sources.  

This work follows SPS97 in its main objectives. A larger
statistics of GRBs is used, and, most 
importantly, the procedure of discriminating between simple and complex bursts 
is formalized and made more efficient.
Due to the latter improvement, the direct test for a complexity - peak brightness 
correlation gives a very significant and meaningful result, which is described in
\S~3. Then, in \S~4, we describe the results of our studies of the ATP shape
as a function of brightness. These studies show further effects  which  are 
also associated  with the complexity - brightness correlation. In \S~5, we 
present a model example demonstrating how the above correlations could occur
using simple assumptions and discuss the issue of the GRB intrinsic luminosity 
function. 
In the Appendices, we present a detailed description of the data 
analysis for the ATP study as well as a number of additional tests for possible 
systematic errors. 

Some of the results presented here were also presented in a preliminary form 
in Stern, Svensson, and Poutanen (1997).

When using words ``brightness'', ``bright'', ``dim'', ``strong'', or ``weak'', 
we always refer to the {\it apparent} peak brightness (peak flux or peak count rate), 
except in those cases when we clearly write {\it intrinsic}. 

\section{Data Analysis} 

This work is based on the data  in the  publically available Compton Gamma-Ray 
Observatory data archive at Goddard Space Flight Center. 
Our sample includes bursts up to trigger number 6230 and contains 1395
events selected as useful for the complexity -- brightness analysis and 1310 
events being useful for the ATP study. We use 0.064~s and 1.024~s time 
resolution data from the Large Area Detectors (LAD).  All the time profiles are 
constructed 
in 64 ms resolution with a 1024 ms resolution extension if necessary. All 
background fits are done using the 1024 ms data as they cover a wider time range
including the pre-trigger history. We use the count rate summed over 
the four LAD energy channels
covering the 25 -- 1000 keV energy range, when studying the behavior of the 
ATP, as well as the count rate summed over channels 2--3
(50 -- 300 keV) in our complexity -- brightness analysis (see SPS97). 
Backgrounds were subtracted using linear fits. A visual scan of all triggered 
bursts was performed in order to select useful events and to set the fitting 
windows.

As a measure of peak brightness we use the peak flux, $F_p$, or 
the peak count rate, $P$, depending on the application. Peak fluxes in 64 ms time 
resolution from the current BATSE catalog (Meegan et al. 1998) were used
when sorting bursts into brightness groups: this is a traditional
measure which is more convenient when comparing results of different works.
On the other hand, peak count rates have a more direct association  with 
the trigger efficiency. Therefore, 
in \S~3, we use our estimates of the 64 ms peak count rate as a measure of 
brightness instead of using the peak fluxes from the catalog. In order to reduce 
brightness dependent biases associated with 
Poisson noise, we developed a peak search scheme  where  each count rate 
excess over 
neighboring time intervals is tested for its statistical significance.

Important procedures of the data analysis and their tests are 
described in Appendix A. 

\section{Different Brightness Distributions for Simple and Complex Bursts }

It is comparatively easy to distinguish between
complex and simple events for the bright GRBs. 
One can then introduce a numerical 
measure of complexity, e.g., the total length of up and down variations 
normalized to the peak count rate. Alternatively, one can count runs
up and down and use their number to characterize complexity (Lestrade 1994). 
Unfortunately, such an approach fails when dealing with weak GRBs because of
the Poisson noise which heavily dominates any of the measures mentioned above
(various attempts to filter out the noise do not help).

The only possibility to extract a ``simple'' subpopulation from the weak burst
population
is to use the ``canonical'' shape of a single pulse, which is more or less 
identifiable by the human eye or by a $\chi^2$ fit. The former approach was 
used by SPS97 in the form of a visual blind test. All events were rescaled to 
the same (low) brightness adding proper Poisson noise. Then each rescaled 
event was
classified by three test persons as ``simple'' (a single FRED pulse), or as 
``complex'' (not a FRED pulse),  
or ``unresolved'' (usually too short to be confidently
identified). Then the peak count rate distributions for simple and complex GRBs 
were compared. Two of three test persons showed that complex GRBs are 
systematically brighter at a significance level exceeding 0.01, the third
test person showed the same effect at a significance level of 0.1. 
We now present results using the $\chi^2$ tests on a larger statistics of GRBs.

\subsection {The $\chi^2$ Separation Between Simple and Complex Bursts}

Unlike the visual test of SPS97, the $\chi^2$ test is aimed at extracting 
events {\it dominated} by a single FRED pulse rather than consisting of a 
single FRED pulse. That is, the main peak was checked for how well 
it is fitted by the FRED's pulse shape, while
other smaller peaks not affecting the fit were allowed for an event to be 
qualified as ``simple''.
Pulses were fitted with the parameterization of Norris et al. (1996):
$C_{r,d}(t) = C_p\exp(-[|t-t_{\max}|/\taurd]^\nu)$, where $\nu$ 
was allowed to vary between 0.9 and 2.2, $t_{\max}$ to vary within $\pm$ 
3 seconds around a direct estimate of the peak position. The
three parameters, the peak count rate, 
$C_p$, the rise time, $\taur$, and the decay time, $\taud$, were free. 
The fitting time interval was 
($t_{\max}-6\taur^{1/\nu}, t_{\max}+6\taur^{1/\nu}$).
The fit for each event was performed for 64, 128,
256, and 512 ms bins. The maximum value of $\chi^2$ per degree
of freedom ($\chi^2_m$) for these four 
variants of binning was used for the classification.
 All events with a peak count rate $>$ 55 counts/64 ms were tested
(dimmer events were discarded), and all of 
them were rescaled to a peak count rate (in energy channels 2+3) 
of 55 counts/64 ms. All events 
with $\taur + \taud <$ 2 s were discarded as the time resolution for short peaks 
at this small brightness is insufficient.

A visual examination of the fitting procedure showed that it is not perfect, but
this is natural. Sometimes a bright GRB which looks as an apparently complex
dense bunch of pulses gave a good $\chi^2$ when rescaled to low brightness, and 
sometimes a GRB looking like a FRED had a bad $\chi^2$ not satisfying the 
parameterization (in such cases no human intervention occurred, of course).
Nevertheless, in 90\% of the events the result of the $\chi^2$ test coincided 
with the visual 
impression, so we conclude that the above parameterization and the whole 
procedure work satisfactorily. Typically, the largest $\chi^2_m$ was 
obtained for the widest binning, 512 ms. 

\begin{figure}
\centerline{\epsfig{file=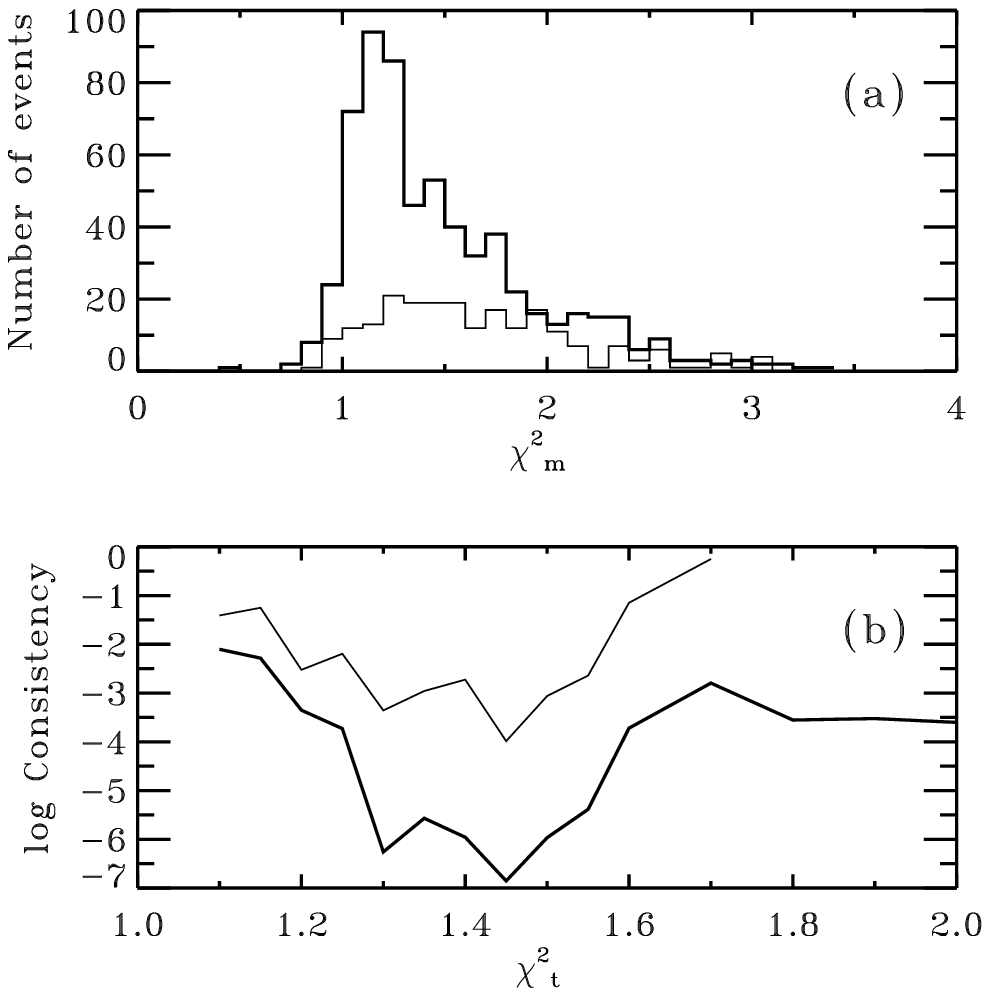,width=9.2cm} }
\caption{a) The reduced $\chi^2$ distribution for the 852 ``resolved'' events with 
$\taur + \taud >$ 2 s. Thick histogram: GRBs (626 events) with faster rise than decay 
($\taur < \taud$). Thin histogram: GRBs (226 events) 
with slower rise than decay ($\taur > \taud$). 
b) The consistency level using the Kolmogorov-Smirnov criterion for the peak count 
rate distributions of simple ($\chi^2_m < \chi^2_t$) and complex ($\chi^2_m > \chi^2_t$) 
GRBs as a function of the $\chi^2_m$ threshold $\chi^2_t$.
Thick curve: the condition $\taur < \taud$ is applied when distinguishing 
simple and
complex events. Thin curve: the condition $\taur < \taud$ is ignored.}
\label{fig1}
\end{figure}

The number of resolved events for which the classification simple-complex
was possible is 852 (of 1395 useful events). 
The distribution of $\chi^2_m$ for these events is presented in 
Figure~\ref{fig1}a. Note the striking 
difference in distributions for ``fast risers'' and ``slow risers'' which 
confirms that the test finds many really simple events among fast risers 
(which can be simple bursts) and almost none of them among slow risers
(which should be complex bursts). 

We found an apparent positive correlation between $\chi^2_m$ and the peak count 
rate (the reason why we use the peak count rate instead of the peak flux
as the measure for brightness is discussed below). We separated bursts by
different thresholds $\chi^2_t$ in $\chi^2_m$ and compared peak count rate 
distributions for simple ($\chi^2_m < \chi^2_t$) and complex ($\chi^2_m > \chi^2_t$) 
events estimating their
consistency levels with the Kolmogorov-Smirnov (KS) test. The results are 
presented in Figure~\ref{fig1}b. We see that 
without applying the  $\taud > \taur$ criterion we have a significance level for
the complexity -- brightness correlation of $\sim 3 \times 10^{-4}$ (i.e, a disproof of
the null hypothesis that the brightness distributions for simple and complex 
bursts are the same), while applying the  $\taud > \taur$ criterion 
the significance level is $< 10^{-6}$ (with the extermal value of 
$1.4 \times 10^{-7}$). 
Our final choice for the classification of a burst as "simple" is: $\taud > \taur$,
$\chi^2_m < 1.45$ (all other events satisfying $\taur+\taud>2$ s are 
classified as complex). 

\subsection{Peak Count Rate Distributions for Simple and Complex GRBs}

\begin{figure}
\centerline{\epsfig{file=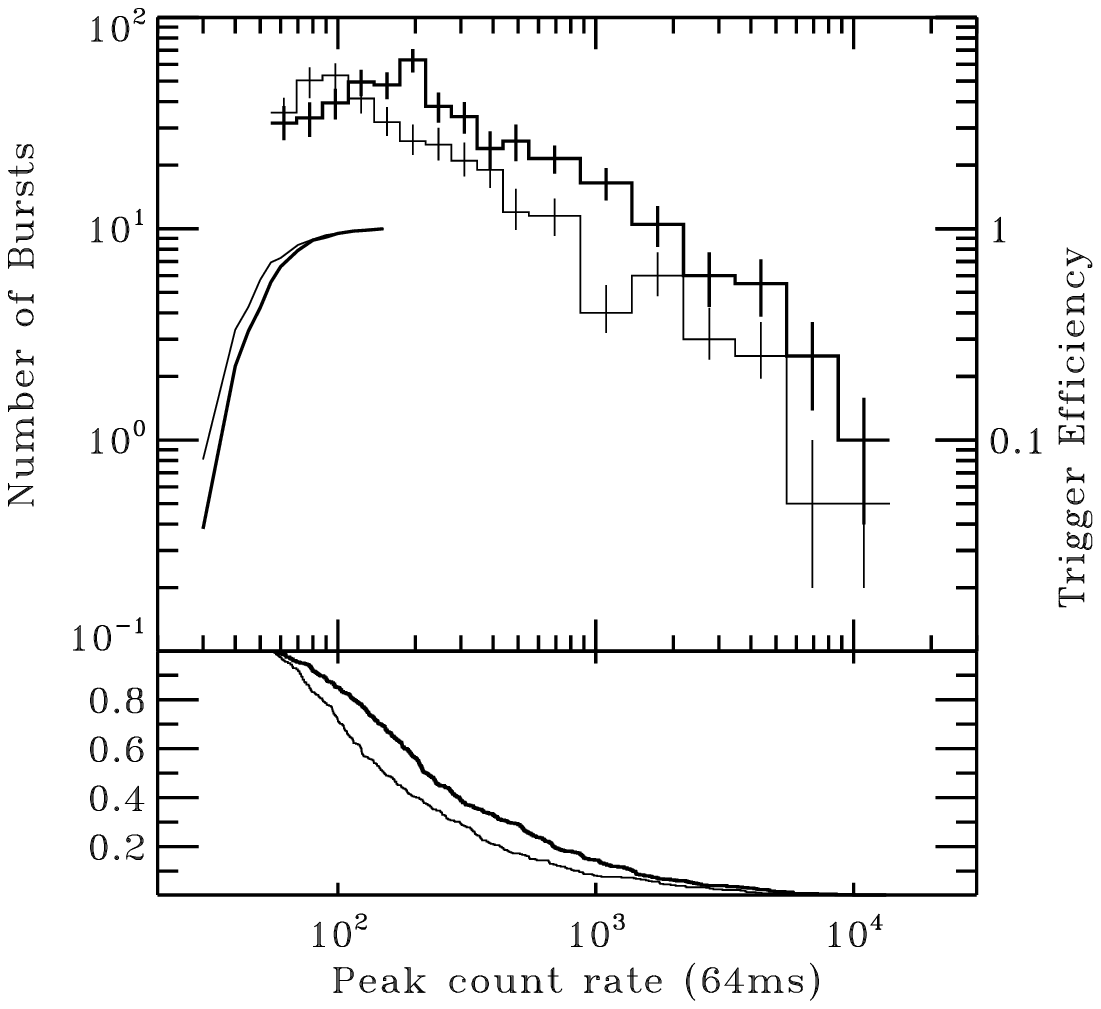,width=9.0cm} }
\caption{Differential peak count rate distributions for simple (thin histogram) 
and complex GRBs (thick histogram) 
corrected for the trigger efficiency (upper panel). 
The trigger efficiency for simple (thin curve) and 
complex (thick curve) bursts are plotted in the lower left corner.  
Lower panel gives the normalized cumulative distributions for simple GRBs 
(355 events) and 
complex bursts (497 events) used in the Kolmogorov-Smirnov test. }
\label{fig2}
\end{figure}

We construct differential peak count rate distributions ($\log N - \log P$) 
for simple and complex events separately.  These 
distributions corrected for 
the trigger efficiency (each class has its own efficiency) are presented in 
Figure~\ref{fig2}.
We plot the number of bursts as a function of peak count rate, while the 
values for the peak flux from the current BATSE catalog (Meegan et al. 1998) 
corrected for the spacecraft orientation, number of triggered detectors, and 
reflection from the atmosphere
could be a better measure of brightness. The effect in the BATSE catalog 
peak flux distributions of 
simple and complex GRBs is similar: 
the KS test gives a consistency level of $10^{-5}$ when using 1024 ms 
peak fluxes and $10^{-7}$ for 64 ms peak fluxes.  
Unfortunately, it is much more difficult to estimate the trigger efficiency
as a function of the peak flux.  The trigger efficiency is a much
sharper and better defined function of the peak count rate than of the 
peak flux. This simplifies the problem of correcting for the trigger 
efficiency when using  peak count rate units.

Besides the trigger efficiency there exists a bias for the brightness dependent dead 
time  for burst read-outs. (The trigger during read-out time is revised so that 
a new event brighter than the triggered event on a 64 ms time scale overwrites 
the first trigger while a weaker event will be lost.) The effect can be accounted 
for by discarding from the sample all events which overwrite a previous trigger. 
We found that the effect is small enough (only 5\% of our events are ``overwrites''). 
Moreover, the effect is almost the same for simple and complex events (within the
statistical accuracy) and does not affect the difference in their distributions. 
In order to save the statistics, we work with the full sample of useful bursts. 

The difference in the behavior of the $\log N - \log P$ distributions for the 
two classes
of GRBs is striking. While the curve for simple events is consistent with a 
power law down to the trigger threshold, the complex class demonstrates an 
apparent break at a count rate exceeding the threshold by a factor of 3 -- 5. 
If one shifts the ``complex'' distribution by a factor $1/k <1$ discarding 
bursts moving below the threshold (i.e., $P/k <$ 55 counts/64 ms) then the KS 
consistency level peaks at $k=2.1$ which can be interpreted as complex events 
being $\sim 2$ times brighter on average. This factor can be even larger 
depending on the behavior of the ``simple'' distribution below threshold. 

 The main difference in the count rate distributions is concentrated in the low 
count rate range and can be described as a relative deficit of complex 
bursts by a factor $\sim 2$ near threshold.
The low count rate range may contain systematic biases (threshold 
effects) and  
we must estimate them before drawing any conclusions. 
We have no reason to suspect the rescaling procedure as being a source of bias
as this procedure is trivial and one can exactly account for the variation of 
noise when we rescale bursts to the same brightness. 
Another possible bias can be associated with the errors of the background 
fits, but this bias has the opposite sign of the effect: weak bursts should
give a higher $\chi^2$ due to nonzero residuals outside of the peak.
The visual classification of SPS97,  which is less sensitive to rescaling and
insensitive to the background subtraction, gave a similar result.

We also cannot suspect systematic errors  in the estimates of the peak count 
rates. The reasonable linearity of these estimates is demonstrated in 
Appendix A (see Fig.~\ref{fig8} there). 
A correlation between complexity and the peak count rate estimates  
could exist but this is a 10\% effect, while a factor 2 bias
is required to account for the effect.
A more serious source of systematic error could be a different trigger 
efficiency for simple and complex events. Indeed, slow risers, which have 
lower trigger efficiency, mostly belong to the complex sample. In order to 
verify 
this, we calculated the trigger efficiency separately for simple and complex 
events as described in Appendix~A (the distributions in 
Fig.~\ref{fig2} as well as the significance levels given above are 
already corrected for different trigger efficiencies).

The efficiency for complex events is actually lower, e.g., at a peak count 
rate of 55 counts/64 ms it is 0.69 for simple events and 0.56 for complex 
events. Nevertheless, the difference is negligible compared to that required to
explain the effect: at least a correction by a factor of two applied only 
to complex weak bursts is required to make the distributions consistent. 
We can hardly admit as  explanation that it is  the result of a huge unknown
selective bias which affects only high $\chi^2$ bursts but does not affect
low $\chi^2$ events. We suggest that the upper curve in 
Figure~\ref{fig2}, representing
presumably intrinsically strong events, extends to larger cosmological 
distances, where some effect associated with high redshifts becomes important.
Then, it is natural to suggest that the lower curve will show a similar 
behavior below the threshold. 

Recently, Pendleton et al. (1997) found a similar effect that
separating GRBs into subclasses affected their $\log N - \log P$ distribution.
The presence of a hard tail in the spectrum (significant emission above 300 keV) 
was used as a criterion for separation. It is possible that the  
``no high energy'' and ``high energy'' subclasses of Pendleton et al. 
correlate with our simple and complex groups, respectively.

In a cosmological scenario, 
it is natural to suggest that the intrinsically strong subpopulation of BATSE GRBs
extends to $1 < z < 3$ where the Universe evolves strongly.  
The evolution at this epoch is clearly visible in the QSO redshift 
distribution (Hartwick \& Schade 1990) and 
in the star formation rate curve (Madau et al. 1996; Abraham 1997).  
In the neutron star 
merging scenario (Blinnikov et al. 1984; Paczy\'nski 1992),  
the GRB rate should be associated with the star formation 
rate (Lipunov et al. 1995; Prokhorov, Lipunov, \& Postnov 1997). 
In this case, the break in the $\log N - \log P$ curve for complex 
GRBs should correspond to the peak in the star formation rate at $z \sim 1.5$. 
In the galactic halo model, a cutoff in the neutron star radial distribution 
could also account for  the observed break.  

Another interpretation is that the different  $\log N - \log P$ distributions 
of simple and complex bursts could be associated with a possible existence of 
two different classes of events among the simple GRBs. 
One class could have a non-cosmological 
origin and therefore Euclidian $\log N - \log P$ distribution which when being 
added to the cosmological component could explain the effect. 
Extending the $\log N - \log P$ distribution using non-triggered bursts could 
provide the answer. 

\section{Brightness Dependent Correlations in  the Average Time Profile}

After confidently observing a complexity -- brightness correlation, 
the ATP -- brightness correlation can be considered to be a direct consequence of 
the former. Nevertheless, the later still provides an independent confirmation of the
correlation tendency, complete the picture, and is interesting in itself.  
For this reason, we present the latest results  of our ATP  studies together
with a more detailed description than was given in SPS97.  

\subsection{On the Stretched Exponential Shape of the ATP}

\begin{figure}
\centerline{\epsfig{file=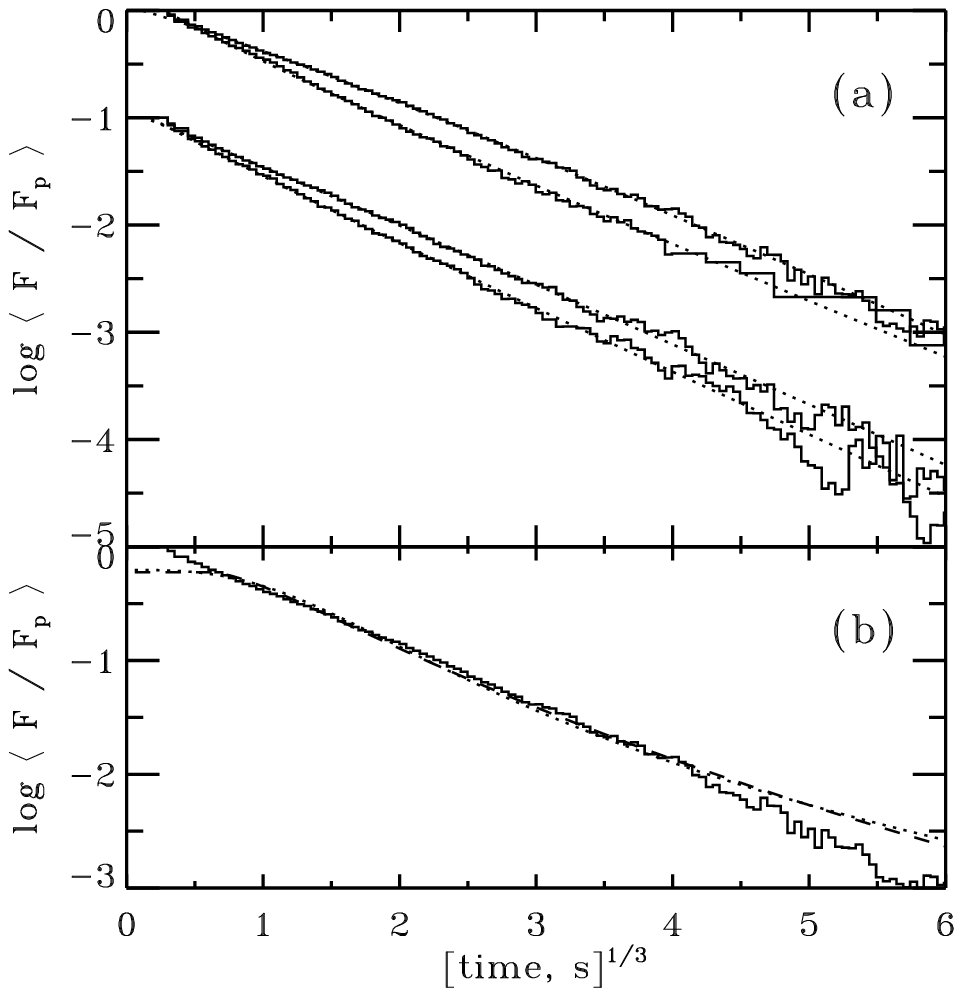,width=9.2cm} }
\caption{a) Rising and decaying slopes of the ATP with their stretched 
exponential fits. Upper set of curves: the full useful sample, 1310 GRBs; 
lower set of curves (shifted down for clarity): the 953 brightest GRBs. 
Rising slopes are steeper than the decaying for both cases.
Dotted curves represent the best 3-parameter ($\beta, \nu, \t0$) stretched 
exponential fit. Best fit values of $\nu$ for the whole sample are
0.30, 0.37 for the rising and decaying  slopes, respectively. 
For the bright sample, $\nu\approx 0.33$, for both slopes. 
b) Examples of fits of the decaying ATP  using other functions. 
Dotted curve: the parameterization 
of Mitrofanov et al. (1997), 
$f(t) = \beta [\t0/(\t0+t)]^{\alpha}$; dashed curve: the one-sided log-normal
distribution (see text) }
\label{fig3}
\end{figure}

 A stretched exponential (SE) shape of the ATP, 
$f(t)\equiv <F(t)/F_p>=\beta \exp[-|t/\t0|^{\nu}]$,  
was claimed by Stern (1996) for the decaying slope
and was confirmed with successively higher 
statistics by Stern \& Svensson (1996) and SPS97 (in
the latter work, an SE shape of the rising slope of the ATP was also 
demonstrated). Both the rising and the decaying slopes of the ATP for the 
overall useful BATSE statistics (1310 events) are presented in 
Figure~\ref{fig3}a. The high statistics demonstrates that the picture 
is more complicated
than the idealized assumptions in SPS97. The rising and the decaying parts
have not only different time constants 
but also different shapes (i.e., different SE indices, $\nu$). 
The decaying ATP is
perfectly described by an SE with $\nu = 0.37$, while the rising ATP is 
described 
by an SE with $\nu=0.30$. Note that the whole ``disorder'' comes from weakest 
bursts. If we remove them, both the rising and decaying ATPs are much closer 
to the ``canonical'' shape with $\nu=1/3$ (Fig.~\ref{fig3}a). 
The different ATP shape of the weakest bursts was noted by SPS97. 
The difference is significant and cannot result from the trigger selection 
effect.
The interpretation of this fact is discussed below. For now we just note that:

- An SE shape with $\nu=1/3$ for both slopes is still a good working hypothesis 
for stronger bursts. 

- For the full sample, the SE indices of the rising and decaying ATPs differ. 
The difference is real and significant ($\sim 4 \sigma$ effect). 
However, both ATPs are still described by SEs. We demonstrate 
below that the difference in ATP shapes is a natural consequence 
of the asymmetric shape of elementary pulses. 


Let us now consider the issue of whether 
the stretched exponential shape of the ATP is just a successful fitting 
hypothesis among other comparable 
possibilities or if it is a natural intrinsic feature of the ATP?

The $\log f$ - $\log t$ plot of the ATP does not 
resemble a power law in any time interval. 
Nevertheless, Mitrofanov, Litvak \& Ushakov (1997) used a
fitting expression with power law asymptotic, $f(t) = [\t0/(\t0+t)]^{\alpha}$,
to fit the ATP. 
Introducing a third parameter, the general multiplier $\beta$ 
(i.e.,  $f(t) = \beta [\t0/(\t0+t)]^{\alpha}$), the best fit
for the decaying part of the ATP 
in the 0.125 - 216 seconds range gives $t_0=3.62$, $\alpha=1.33$ 
with $\chi^2 = 128$ for 19 degrees of freedom
(Fig.~\ref{fig3}b) and with an unreasonably small value at $t=0$: 
$f(0) = \beta=0.62$. 
Since Mitrofanov et al. (1997) constructed the ATP in 1024 ms resolution (i.e., 
the $t < 1$ s region was lost) and only within the $t < 20$ s range, it is not 
surprising that they obtained a good fit in this narrow interval. However,
for a wider time interval this parameterization is unacceptable.

Let us take as another example the log-normal distribution which is quite 
common in Nature. It has the wrong asymptotic at $t = 0$, but we correct this
by setting $f(t)=\beta \exp(-\log^2(t/\t0)/2\sigma^2)$ at $t > \t0$ and 
$f(t)=\beta$ at $t < \t0$. 
This semi-artificial expression fits the ATP in the same range as above with
$\chi^2 = 89$ for 19 degrees of freedom (best fit parameters $t_0=0.20$ and 
$\sigma=3.18$) and again with an unsatisfactory value 
at $t=0$: $f(0) = \beta = 0.6$ (see Fig.~\ref{fig3}b).  
For comparison, an SE fit of the same ATP gives $\chi^2
= 4.1$ at $\nu =0.37$, $\t0 = 1.12\s$ , and $\beta =1.04$. For the reasonableness 
of the $\chi^2$ values, see Appendix~B.

\begin{table*}[htbp]
\caption{Time constants of the averaged time profiles (ATPs) }
\begin{center}
\begin{tabular}{l l l l l l l l}
\tableline
\# & Peak flux &  $N$ &  $\tr$ & $\td$ &  $\tr+\td$ &  $\td/\tr$ & $\chi^2$\\
\tableline
1 & 12 -- $\infty$  & 84  & 0.29$\pm$0.08 & 0.36$\pm$0.07 & 0.65$\pm$0.13 
& 1.23 $\pm$0.17 & 7.6 \\
2 & 3 -- 12         & 282 & 0.35$\pm$0.04 & 0.51$\pm$0.06 & 0.87$\pm$0.09 
& 1.47 $\pm$0.11 & 6.9 \\
3 & 1.75 -- 3       & 239 & 0.37$\pm$0.05 & 0.64$\pm$0.08 & 1.01$\pm$0.12 
& 1.74 $\pm$0.21 & 4.8 \\
4 & 1 -- 1.75       & 358 & 0.45$\pm$0.05 & 0.73$\pm$0.07 & 1.17$\pm$0.11 
& 1.63 $\pm$0.11 & 21 \\
5 & .7 -- 1         & 196 & 0.43$\pm$0.06 & 0.74$\pm$0.10 & 1.17$\pm$0.15 
& 1.70 $\pm$0.15 & 22\\
6 & 0 -- .7         & 151 & 0.38$\pm$0.06 & 0.77$\pm$0.11 & 1.15$\pm$0.17 
& 2.00 $\pm$0.20 & 29\\
\hline
7 & 7.5 -- $\infty$ & 145 & 0.29$\pm$0.05 & 0.38$\pm$0.06 & 0.67$\pm$0.10 
& 1.33 $\pm$0.14 & 11 \\ 
8 & 5 -- $\infty$   & 209 & 0.29$\pm$0.04 & 0.41$\pm$0.05 & 0.71$\pm$0.09 
& 1.40 $\pm$0.12 & 9 \\
9 & .8 -- 2.5       & 659 & 0.43$\pm$0.04 & 0.72$\pm$0.06 & 1.15$\pm$0.09 
& 1.68 $\pm$0.09 & 25 \\
\tableline
\end{tabular}
\end{center}
\tablecomments{
Time constants, $\tr$ and $\td$ ($\s$),  are given for 
the stretched exponential simultaneous fit with $\nu=1/3$ to the 
pre-peak and post-peak average time profiles, respectively. Peak fluxes in 
64 ms resolution 
($\ph\cminvsq\secinv$) are taken from the BATSE database and are measured 
in channels 2 and 3. 
$N$ is the number of bursts in the given brightness interval. 
The fitting time interval is $0.5 < | t^{1/3} | < 5$.
The errors and the $\chi^2$ values (for 33 formal degrees of freedom) are
estimated as described in Appendix B. 
}
\label{tab1} 
\end{table*}

Probably, there is no 3-parameter  expression which would fit the ATP 
on such a broad time interval 
except for the SE. Note, that our SE fit is in fact a ``2.5-parameter'' fit, 
i.e., the third parameter, the multiplier $\beta$, just accounts for the 
uncertainty associated with the first 64 ms bin of the ATP assumed 
to be close to 1. 
The best fit value of $\beta$ is actually always close to 1 and 
therefore we do not give the $\beta$-values. 

Summarizing the issue, we state that the stretched exponentiality of the 
ATP is probably not exact. However, with the existing statistics we do not see 
any
statistically significant deviations. The SE is a natural distribution shape 
in very different classes of physical phenomena when a wide range of time scales
are involved (Ching 1991; Jensen, Paladin, \& Vulpiani 1992). 
Therefore we believe that our  
choice of SE as the fitting expression is not only efficient but also 
meaningful.   

The procedures of measurement of the SE  time constants of the ATP 
and estimates of statistical errors follow SPS97. They are 
described in more detail in Appendix B.

\subsection{Time Constants of the ATP as a Function of Peak Flux} 

\begin{figure}
\centerline{\epsfig{file=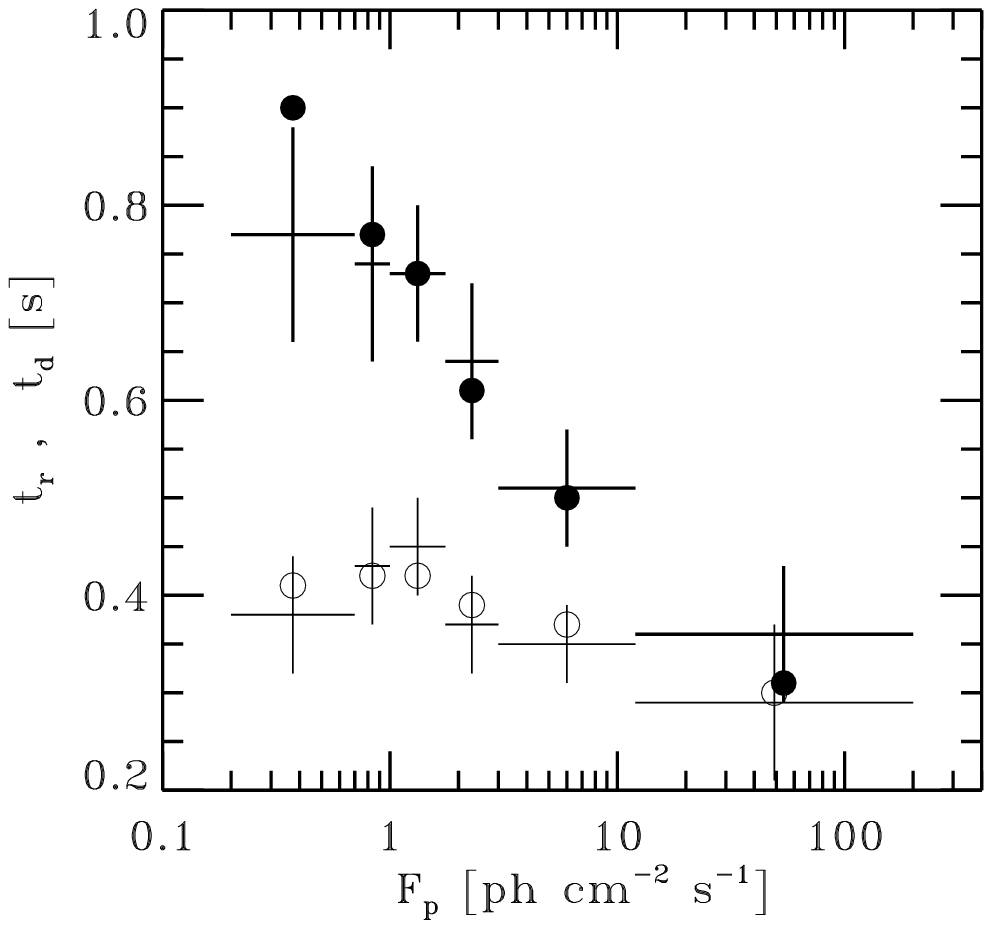,width=9.2cm} }
\caption{Time constants, $t_{r,d}$ (s),  vs peak photon flux in $64 \ms$ 
time resolution, $F_p$ ($\ph\cminvsq\secinv$). The SE index, $\nu$, was fixed 
to $1/3$. 
Upper and lower crosses represent results 
for the post-peak (decaying) and pre-peak (rising) 
ATPs, respectively. Error bars in $t_{r,d}$ correspond
to the values in Table~\ref{tab5} in Appendix~B. 
Error bars in photon flux represent
the width of the brightness group. 
Open and filled circles represent best fit values for $t_{r,d}$ with $\nu$ 
fixed at 0.30 and 0.37 and rescaled with 
the factors 1/0.51 and 1/1.6 for the rising and decaying slopes, respectively. }
\label{fig4}
\end{figure}

We use the following parameterization of the ATP: 
$f(t)=\beta \exp\left[ - \left| t/\trd \right|^{\nu_{r,d}} \right]$, where 
$\trd$ and $\nu_{r,d}$ are the time constants and the SE indices of the 
rising and  decaying ATP, respectively. 
In our studies of the ATP behavior, we use two kinds of SE fits. The first 
is the simultaneous fit of both ATP slopes with SEs of the same index, 
$\nu=1/3$,
and a common normalization factor $\beta$. This 3-parameter fit ($\tr, \td$, 
and $\beta$) has the best statistical accuracy.  
The results of these SE fits are summarized in 
Table~\ref{tab1} and in Figure~\ref{fig4}.


\begin{table}[htpb]
\caption{Time constants measured with $\nu=1/3$ in different energy channels} 
\begin{center}
\begin{tabular}{clll}
\hline
Channel \# &  $\tr$ &  $\td$ & $\td/\tr$  \\
\tableline
1  &     0.50  &   0.70 &   1.40 \\
2  &     0.42  &   0.56 &   1.33 \\
3  &     0.35  &   0.42 &   1.20 \\
4  &     0.20  &   0.27 &   1.35 \\
\tableline
1 - 4  & 0.39  &   0.53 &   1.35 \\
\tableline
\end{tabular}
\end{center}
\tablecomments{
Time constants, $\tr$ and $\td$,  of the averaged time profiles 
are given for the 280 brightest events in separate LAD energy channels. 
The value of the index $\nu$ was fixed at $1/3$.
Relative errors for $\tr$ and $\td$ are $12\%$,
for $\td/\tr$ they are 8\%. 
}
\label{tab2} 
\end{table}

In comparison to the GRBs statistics used by SPS97 (912 events), 
the temporal stretching of the decaying slope for weak events is 
now more pronounced and significant, 
$t_{d,9}/t_{d,7}=1.89^{+0.68}_{-0.44}$ 
(90\% confidence interval) comparing samples 9 and 7  (see Table~\ref{tab1}).
The corresponding rejection level for a null hypothesis (i.e., no temporal stretching) 
has now increased to 0.99995 (assuming a normal 
distribution for the deviations of the values).
The temporal stretching of the rising slope also increased slightly, but this
is still at a low significance level.
The resulting variation of the asymmetry (the ratio $\td/\tr$) slightly 
decreased.
Comparing samples 1 and 9 we get: 
$(\td/\tr)_{\rm dim}/(\td/\tr)_{\rm bright} = 1.48^{+0.55}_{-0.32}$ 
($90\%$ confidence interval). The significance level for this correlation
remains the same as for the SPS97 sample, $\sim 0.02$.
The variation of $\tr+\td$ with peak flux 
has a statistical significance of $4\times 10^{-4}$; 
the ``dim/bright'' ratio becomes 
$(\tr+\td)_9/(\tr+\td)_7=1.71^{+0.62}_{-0.38}$ (sample 9 compared to 
sample 7).

The results presented so far were obtained with a fixed SE index, $\nu=1/3$. 
We now present a second fitting variant. As was mentioned above,  
the ATP for the whole sample indicates a larger $\nu$ for the decaying 
slope and a lower $\nu$ for the rising slope. Despite the fact that the 
difference results from the contribution of weak GRBs, it would be interesting 
to check the behavior of the time constants when $\nu$ is fixed to different values
for the two slopes. Unfortunately, we cannot then use a common $\beta$ for both 
slopes, as 
this would lead to unacceptable values of $\chi^2$. Therefore,  we 
performed {\it independent} 2-parameter fits ($\beta_r, t_{r}$)  and 
($\beta_d, t_{d}$) 
for the two slopes  setting $\nu_r = 0.30$, $\nu_d = 0.37$. 
(These fits give 20\% larger statistical errors for the time constants as 
compared to the simultaneous 3-parameter fit.) 
With such a fit, 
$\tr$ becomes systematically smaller (by a factor 0.51), and  
$\td$ systematically larger (by a  factor 1.6). 
Such an effect on the time constants is due to the strong correlation 
$\nu-t_{r,d}$ 
(see Fig.~\ref{fig10} in Appendix~B).     
These time constants rescaled by a factor 1/0.51 and 1/1.6 are 
shown in Figure~\ref{fig4}. 
One sees that the behaviors of these time constants with peak 
flux are the same. The variation 
of the asymmetry, $\td/\tr$, is even slightly larger than for the first fitting 
variant.

To test for spectral redshift of weak bursts as being the possible explanation 
of the asymmetry variation,  we measured the asymmetry in separate LAD's energy 
channels.  If GRBs were more asymmetric at higher energies, then the 
spectral 
redshift of the softer, more symmetric, component to energies below the 
observational band could cause an asymmetry -- brightness correlation.
The results are presented in Table~\ref{tab2}. One sees no tendency 
supporting such a hypothesis: GRBs have the same asymmetry in all channels. 


A test for the presence of brightness 
dependent biases in these results is described in 
Appendix~A. Only the weakest sample 6 is affected by the trigger selection bias 
and
by Poisson noise. For  this reason, sample 6 was excluded from the analysis of 
the brightness dependent correlations. 
 
\subsection{Variations of the Stretched Exponential Index with Peak Flux} 

\begin{figure}
\centerline{\epsfig{file=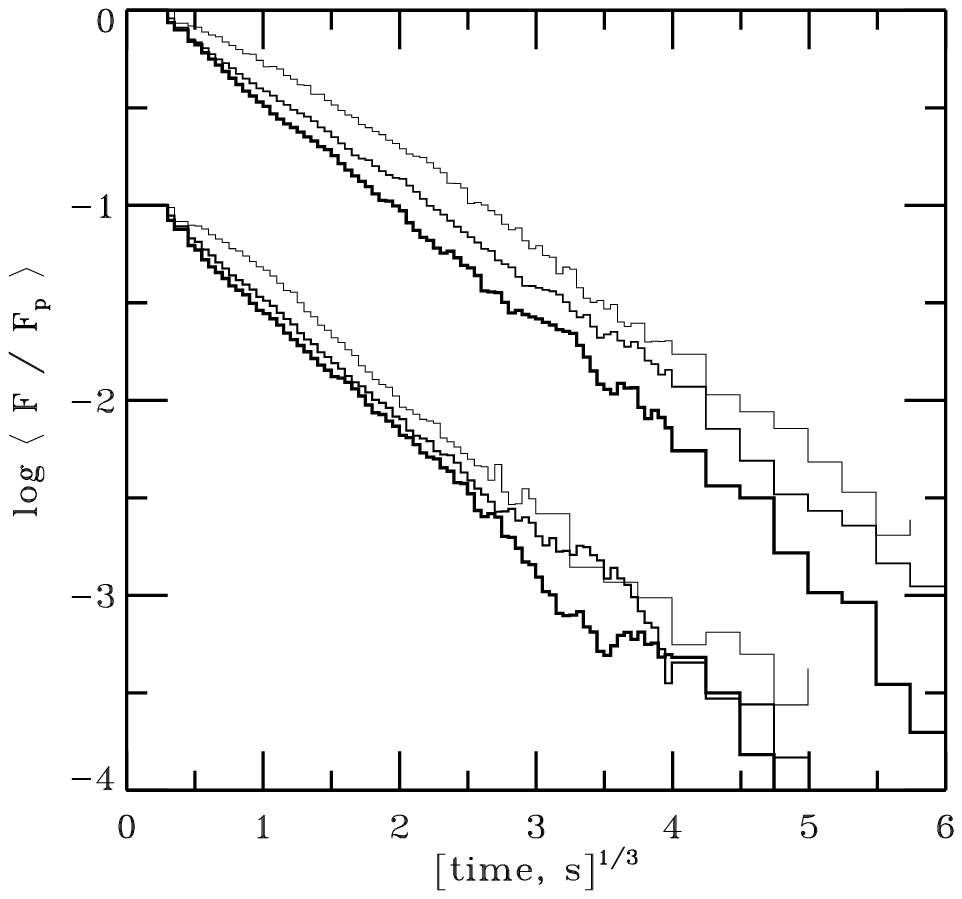,width=9.2cm} }
\caption{Decaying and rising 
slopes of the ATP for different brightness groups: strong (samples 1+2),
medium (samples 3+4), and weak (samples 5+6). Rising slopes are shifted 
down for clarity. Stronger bursts have smaller time constants (slopes are
steeper). 
}
\label{fig5}
\end{figure}

Figure~\ref{fig5} 
presents the decaying and rising slopes for samples 1+2, 3+4, and 5+6. 
The stretched exponential indices for these samples 
are given in Table~\ref{tab3}. 
Except for the apparent ``time dilation'' effect, the deformation of the weakest 
ATP is 
clearly visible. Partially, this deformation results from brightness dependent 
biases (short weak bursts have smaller trigger efficiency and  the 
peak searching scheme is unable  to find the sharp peaks of longer events, 
also see Fig.~\ref{fig9} in Appendix A).
These biases suppress the ATP within 1 s from the peak while its effect beyond 
1 s is not so strong.  


\begin{table}[htbp]
\caption{SE indices for three wide brightness groups} 
\begin{center}
\begin{tabular}{lllll}
\tableline
\#  & Peak flux &  N   &   $\nu_r$       &    $\nu_d$ \\ 
\tableline
1+2 & 3 -- 200  & 366  & 0.338$\pm$0.027 & 0.322$\pm$0.026 \\ 
3+4 & 1 -- 3    & 597  & 0.316$\pm$0.020 & 0.359$\pm$0.023 \\ 
5+6 & 0 -- 1    & 347  & 0.252$^*$       & 0.411$\pm$0.034 \\
\tableline
All & 0-200     & 1310 & 0.300$\pm$0.017 & 0.371$\pm$0.021 \\ 
\tableline
\end{tabular}
\end{center}
\tablecomments{$\nu_r$ , $\nu_d$ are best fit SE indices for 
the 3-parameter (free $\beta, \nu_{r,d}, t_{r,d}$) fit   in the 
$0.5 < |t^{1/3}| < 6$  interval.  \\
$^*$ For sample 5+6, the fit was done in the $1 < |t^{1/3}| <6$ time interval
because of too strong ATP deformations within 1 second. Errors were not 
estimated. 
}
\label{tab3} 
\end{table}

The estimate of the significance level for the shape variation is 
difficult because
trigger selection and Poisson biases give deformations of the same sign as 
(but still much smaller than) observed. SPS97 estimated the significance level
of deformation of the decaying slope for the weakest sample as 0.05 after 
correcting for the above biases and possible error of the background 
subtraction.
We do not make such an estimate here, because the deformations of
the ATP must appear as a direct consequence of the much more 
significant complexity -- brightness 
correlation described above.

\section{On the Issue of the Intrinsic Luminosity Function}

When separating GRBs into brightness groups using their peak fluxes, we 
assumed that the peak fluxes give us an approximate measure of the distance 
(a standard candle approximation). The fact that we see 
correlations between the shape of the ATP and apparent peak 
brightness tells us that the peak luminosity has a rather 
broad distribution and correlates with the ATP. 
What could serve as a better standard candle?

Some researchers suggest that the total energy fluence is a better standard 
candle
than the peak luminosity (e.g. Petrosian \& Lee 1996) arguing that this 
would be more physical. However, 
if we accept that something like the pulse avalanches of Stern \& Svensson 
(1996)
takes place, then a dispersion in the fluence of a few orders of magnitude
seems natural. The GRB itself emits just a fluctuating fraction of the total 
available energy, and probably there are many events where an observer sees 
no GRB. The pulse avalanche model describes the transmission 
of energy from a main reservoir as a highly unstable near-critical process, 
where in the idealized case of exact criticality and infinite available energy, 
the fluence distribution should become a power law. 

 A better candidate for the standard candle would be 
a single pulse independently of whether it alone constitutes a burst or 
appears as one of the pulses in a complex bursts. 
Again, the 
distribution of the pulse peak amplitudes (fluxes) within one GRB 
seems narrower than the distribution of the pulse fluences 
(for studies of the fluence distribution of pulses in GRBs, see 
Li \& Fenimore 1996) -- otherwise a negative  correlation between 
the pulse duration and its amplitude within one GRB would be visible. 
Does any correlation exist within one event?  The visual impression
from GRBs time profiles is that there is no evident correlation.
Unfortunately such a correlation  is not easy to extract as pulses 
tend to overlap and it is
very easy to take a dense bunch of narrow pulses for a single pulse.

\begin{figure}
\centerline{\epsfig{file=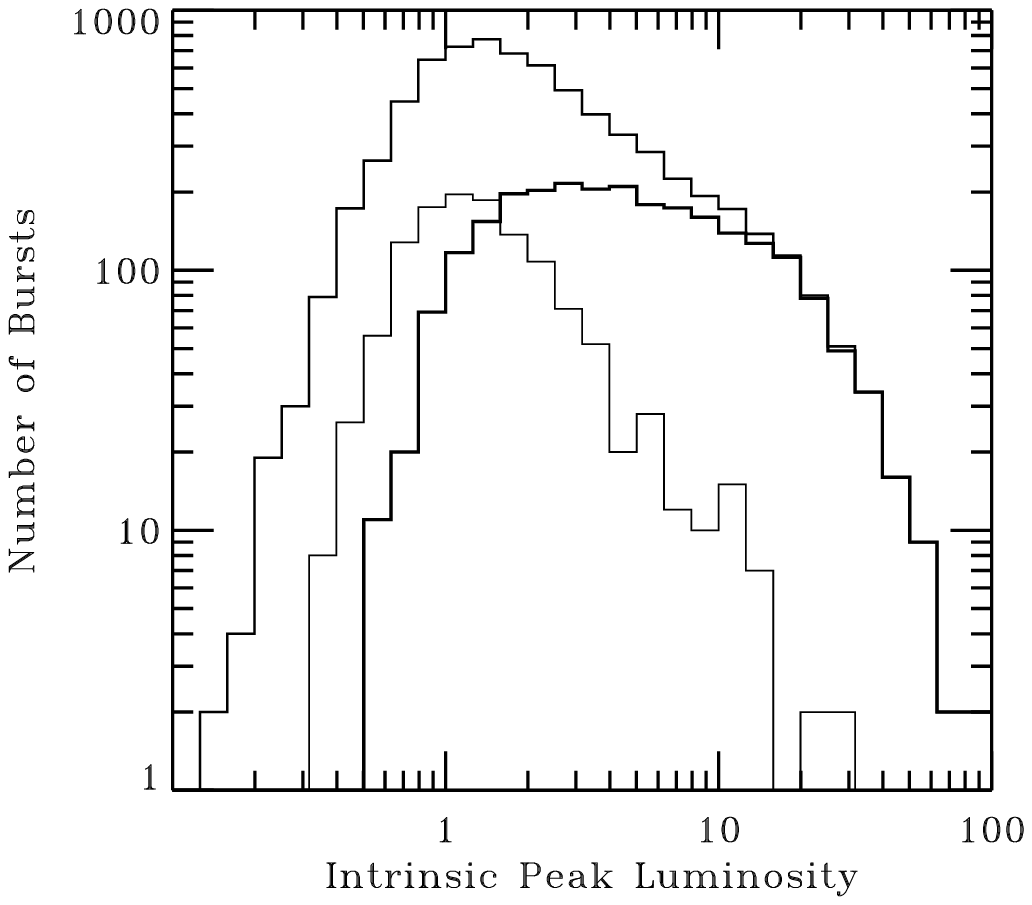,width=9.2cm} }
\caption{Example of intrinsic peak luminosity function for 7000 events simulated with 
the pulse avalanche model (upper curve). The peak amplitudes, $p$, of individual 
pulses were 
sampled with a log-normal distribution, $dn/dp \sim \exp[-\log^2p/2\sigma^2]$ 
with $\sigma = 0.7$.
Lower curves are peak distributions for simple (thin histogram, 1240 events) and 
complex (thick histogram, 2481 events). The separation of simple and complex 
events was obtained using the same procedure as described in \S~3.1  rescaling 
simulated bursts to the same peak count rate, 55 counts/64 ms. 
GRBs with $\taur + \taud <$ 2 s and those that did not pass the trigger 
were not used in the classification procedure. 
The $\chi^2$ distribution for simulated events is more sharply peaked at 
$\chi^2 \sim 1$ than was the case for real
events. 
}
\label{fig6}
\end{figure}

Actually the peak amplitudes should be distributed somehow. Let us consider, as an 
example, a log-normal distribution for the pulse peak amplitude 
(the same distribution was found by Li \& Fenimore 1996 for pulse fluences).
Using this distribution in the pulse avalanche simulations we determined 
the corresponding intrinsic peak luminosity 
function for the simulated bursts caused by the piling up of pulses. 
We then applied the same procedure of $\chi^2$ selection to our simulated bursts
that was used to separate simple and complex real GRBs (see \S 3.1). 
The difference in the distributions of intrinsic peak luminosities of simple and 
complex events in this example (see Fig.~\ref{fig6}) 
seems sufficient to cause a difference in the behavior
of their $\log N - \log P$ curves in the presence of a strong evolutionary 
effect (the ratio of the median peak flux of the distributions is 3.7). 


\section{Summary and Conclusions}

The effects we confidently detect can be summarized as follows:

1. Complex bursts have systematically larger peak count rates than 
simple bursts. Their peak count rate 
distribution has an apparent break at a count rate $\sim 200$ counts/64 ms 
which exceeds the threshold by a factor four. 

2. Weaker bursts have a stronger asymmetry
between the rising  and decaying slopes of the average time profile.
 
3. The weakest bursts have a different shape of the 
average time profile, which can be described as a larger
stretched exponential index (at least for the decaying slope).

4. The general temporal stretching 
of the ATP for weak bursts is complicated by the 
deformations of the ATP resulting from the two previous effects.

Effects 2) and 3)  appear as direct consequences of effect 1).
The effect 4) can be a superposition of both cosmological time dilation 
(\cite{pac92})
and intrinsic luminosity - duration correlation (\cite{bra94}).  
Effects 1) -- 3) can appear only as a result of the correlation 
between temporal properties of GRBs and their intrinsic luminosity (if we 
discard as explanation an evolution of GRBs temporal properties with time which
seems unnatural). The large value of these effects indicates that the intrinsic 
luminosity function is a wide one: comparable with the apparent brightness dispersion
from the distance distribution. Therefore the $\log N - \log P$ curve can 
be a convolution of two functions of comparable dispersions.

Both the most confident and the most informative effect is the first one.
There is no obvious source for such a high near-threshold bias (selective for 
complex events) to account for this effect. There is, however, 
a natural phenomenon that could account for it. As was already
mentioned above, this could be the evolution of the star production rate. 
We must accept this as a possible explanation  if we accept the merging of 
neutron star binaries as the  source of GRBs. 

It seems that we succeeded to extract an intrinsically strong subsample
from the BATSE GRBs which covers distances  to redshifts of $z\sim 1.5$ 
where evolutionary effects 
should  be strong.  A cosmological fit with an evolutionary factor for GRBs
based on the measured star formation rate and with a model luminosity function 
similar to that presented in \S~5 is a matter for separate work.  The crucial
data  to confirm this point of view could appear from the search for 
untriggered bursts in the 
continuous BATSE data records. This search is ongoing (Kommers et al. 1997) 
and  is worth intensifying. 

\acknowledgments

We thank Felix Ryde for helpful assistance. 
This study made use of the data provided by the 
Compton Observatory Science Support Center. 
We are grateful to the BATSE team for a fast supply of new
data to the publically available database. 
This research was supported by grants 
from the Swedish Natural Science Research Council, Stockholm University,
the Swedish Royal Academy of Sciences, the Wenner-Gren Foundation for Scientific 
Research and a NORDITA Nordic Project grant. 

\appendix 
\section{Data Analysis}

\subsection{Background Fitting}
 
The  background fitting was based on a visual scan of all BATSE  bursts. 
This work cannot
be done automatically because there exist many events with complex 
backgrounds contaminated with
non-Poisson features which can mimic the contribution of a GRB.
In fact, each individual burst requires a researcher's decision -- how to fit 
it or whether it should be discarded. To avoid subjective biases 
for weak events, we followed the rules described below: 

1. All fits are linear (the background anyway often demonstrates a 
non-polynomial behavior and a higher order polynomial fit could be unstable).
Linear fits were made over one or two ``fitting windows'', which were set in
``quiet'' time intervals, having a good $\chi^2$. We also set the 
``observational''
windows avoiding background features not associated with the burst.
In the case of a smooth curved background with several burst episodes in the 
event, we
set a few fitting windows including quiet time intervals between burst episodes,
so the background was approximated by a broken line.

2.  Each feature separated by a wide time interval from the main peak
should be tested whether it came from the same direction as the main peak.
To compare directions we made a linear fit to the count rate in each of eight LADs
in the time interval covering the feature.  
We  then maximized the total 
reduced $\chi^2$ in the time interval varying the time resolution, 
and calculated the eight-component vector $\chi^2$. This vector
should approximately be parallel to that of the main peak, otherwise the 
feature should be avoided when setting the observational windows.
The procedure turned out to be an efficient  way to clean up bursts from
unrelated fluctuations. 

3. We adopted a default set of windows: a fitting window ($-$120 s -- $-$70 s,
where the boundaries are given relative  to the trigger time), 
an observational window ($-$70 s -- $+$200 s), and a second fitting window
($+$200 s -- $+$250 s). If possible, we used the default set. 
If the background has a moderate curvature, it should not 
give a bias as it has a random sign. This rule reduces a possible
subjective bias. A narrower time interval between the fitting windows 
was allowed if the 
burst was apparently short
or if the background was strongly curved.

4. All events, where we were unable to make a confident conclusion that
we did not loose and did not contaminate the signal at 100~s after and 
50~s before the highest peak, were
discarded. This could be due to wide data gaps, strong solar flares,
rapid variations of the background, or 
chaotic variations with bad $\chi^2$ from the wrong direction. 

\begin{figure}
\centerline{\epsfxsize=11cm {\epsfbox{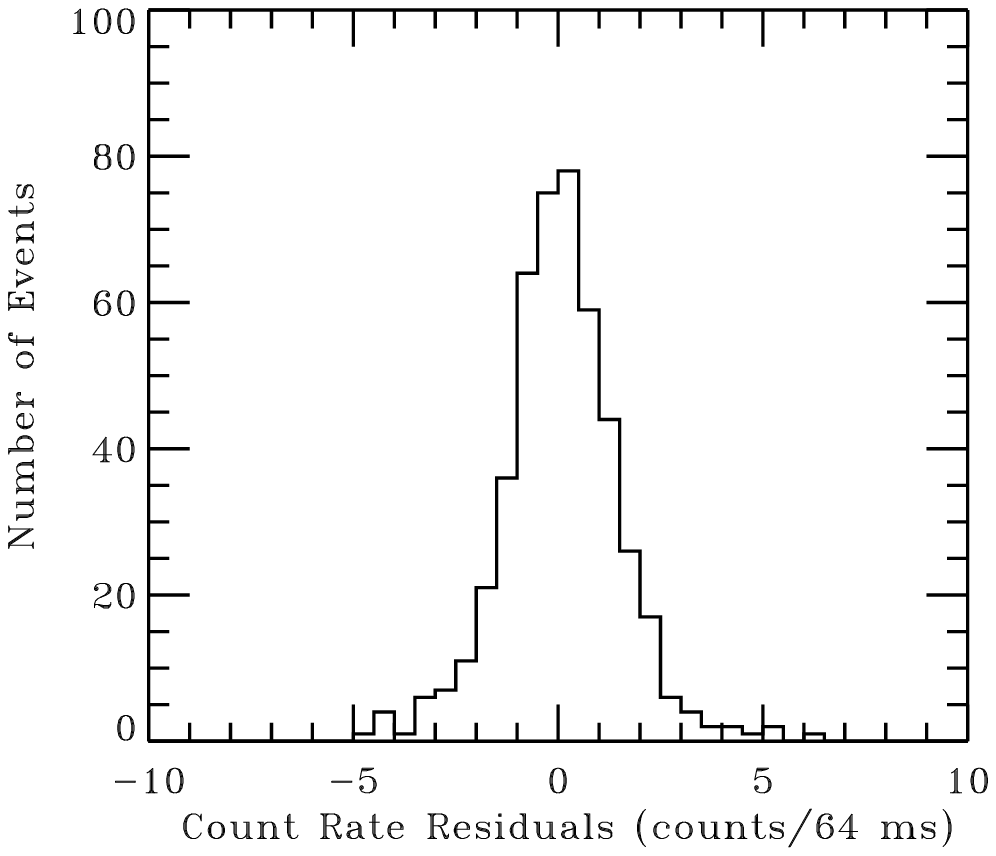}}}
\caption{
Residuals of our background fits in ``testing windows'' located in time 
intervals where the contribution of the GRB is invisible.
The background fits were obtained using ``fitting windows'' (see text).
\label{fig7}} 
\end{figure}

This fitting procedure gives as small bias as possible  when summing up 
signals from different GRBs to get the average time profile, despite the fact 
that it is not accurate when dealing with individual events. 
 To estimate the magnitude of possible errors introduced by the fitting 
procedure we made the following
test. For each event where it was possible ($\sim 75\%$ of all events), 
we selected testing windows between our fitting windows
avoiding regions with GRBs signals and measured the residuals of our fits 
in the testing windows.  The distribution of residuals 
for 469 weak and medium events is shown in Figure~\ref{fig7}. 
The distribution of residuals is reasonably symmetric, the average value for
the residual is +0.084 counts/64 ms, the $1\sigma$ variance is 1.43 counts/64 
ms. This is an argument that we have no significant
systematic error (that exceeds the statistical error). The statistical 
error for the average residual is 0.11 counts/64 ms. For comparison, 
the typical background in channels 2 and 3 is 300 counts/64 ms, 
the peak count rate for the weakest events is $\sim$ 50 counts/64 ms. The error
introduced into the average time profile is inversely proportional to the
peak count rate. For medium and bright events we can neglect the 
fitting uncertainty -- it is much less then $10^{-3}$. The exception is
the weakest group where the $1 \sigma$ error in the relative averaged 
residual is $1.5 \times 10^{-3}$.

\subsection{Selection of the Highest Peak}

The direct peak selection using the  highest 64 ms bin, while working good for
bright events, suffers from Poisson noise for weak events. It just takes
the highest Poisson fluctuation as being the burst's highest peak.
Nevertheless, as far as we adopted the 64 ms resolution approach, we should not 
use a different time resolution for weak bursts. 
As a compromise, we developed a hybrid scheme 
which combines a search for the peak interval in a lower time resolution and 
a search for statistical significant deviations in this interval with a 
higher time resolution. 

First, we determine the shortest time-scale, 
$\Delta t_j = 64 \ms \times 2^j$, $j = 0,...,4$, where a GRB has 
statistically significant variations between neighboring time bins
(a $7 \sigma$ threshold, which corresponds to a $5 \sigma$ threshold 
for a deviation from the average). 
Then we search for the highest peak centered at bin number $k$ using the 
time resolution, $\Delta t_j$, and calculating the peak flux as 
$\max\{\sum_i c(i) \exp[-((k-i)/2^j)^2]\}$ 
where $k$ and $i$ are indices
of $64 \ms$ bins, and $c(i)$ is the count rate  in the $i$-th bin.

Then, if $j\geq 1$ we make a  second iteration: searching for statistically 
significant excess over average within the brightest $\Delta t_j$ interval
using a shorter time-scales. The significance threshold, $h_l$ ($l<j$), is 
reduced at this step. The thresholds were optimized empirically
when tuning the scheme by rescaling strong bursts to the weakest, adding 
corresponding Poisson noise, selecting the highest peak, and comparing
resulting peak amplitude and position with true values. With 
thresholds $h_0 = 4.0 \sigma$, $h_1 = 3.2 \sigma$, $h_2 = 2.0 \sigma$, and
$h_3 = 1.2\sigma$ above the average count rate in $\Delta t_j$
we obtain a reasonable linearity between the expected and the measured count 
rate and still preserve the 64 ms resolution.

\begin{figure}
\centerline{\epsfxsize=13cm {\epsfbox{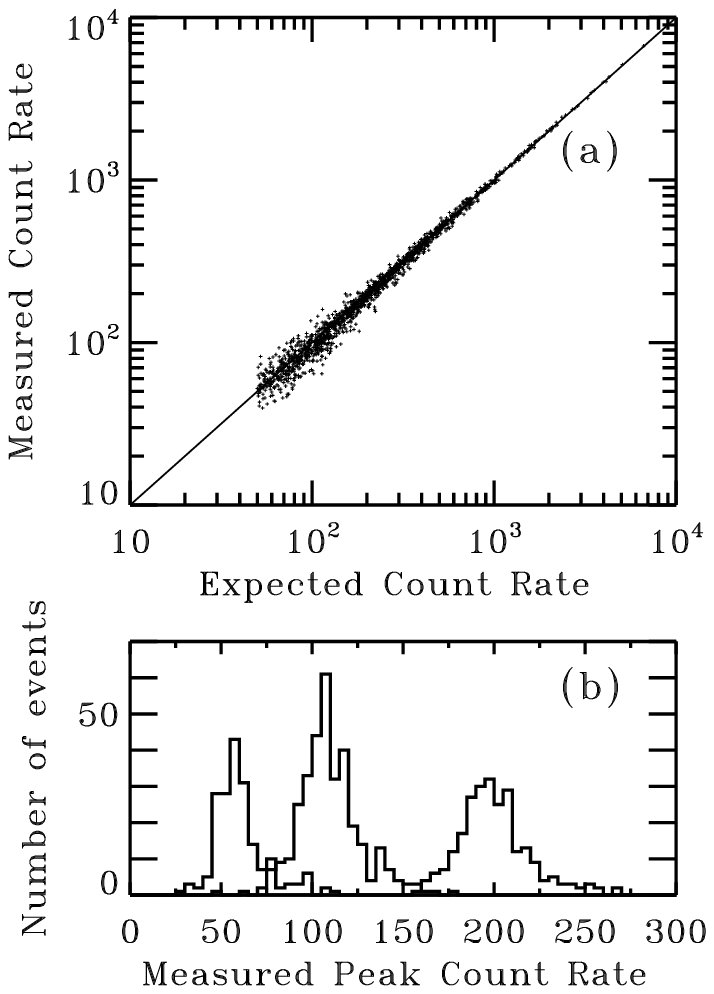}}}
\caption{
Test of the procedure for the peak count rate estimate.
a) expected vs. measured count rate scattering plot obtained with bright events
longer than 2 s rescaled to lower brightness. Note that the peak count rate 
estimate is based on the 64 ms resolution.
b) Distributions of the measured peak count rates at fixed expected count 
rate
$P$: $P$ = 55 counts/64 ms; $P$= 110 counts/64 ms; $P$ = 200 
counts/64 ms. Bursts 
which did not pass the simulated trigger are not included.
For description of rescaling and triggering procedures, see \S~A.3. 
\label{fig8}} 
\end{figure}

A test of this procedure by rescaling strong events to a given value 
of peak count rate is presented in Figure~\ref{fig8}. The 
rescaling procedure is described in \S~A.3. 
Only those events which passed 
our simulated trigger after rescaling were included into the 
test distributions. One can see from Figure~\ref{fig8}a 
that we have a reasonable 
linearity between the measured and the expected peak count rates 
(compare with a similar plot in Fig.~2 of  in 't Zand \& Fenimore 1994).
The systematic bias in the peak count rate 
for the weakest events is within 3\%, the relative error is $\sim 35\%$ (FWHM) 
for $P$= 55 counts/64 ms and $\sim$22\% for $P$=110 counts/64 ms. There are 
non-Gaussian 
tails towards higher values associated with Poisson fluctuations.
They, however,  
do not exceed a few per cent of the peak integral. If one increases the 
thresholds, these tails will disappear, but a nonlinear bias of peaks towards
lower values will appear. The thresholds  have been set so as to 
have a reasonable average linearity of peak count rate estimates for 
weak events. 
(Note that direct selection of
the highest bin in 64 ms resolution systematically overestimates the peak 
amplitude
by a factor of 2 for the weakest events). The errors in the peak position 
can be characterized as follows: in 35\% of the cases, 
the error does not exceed one $64 \ms$ bin, with 47\% probability the error is
within $0.128 \s$, with 83\% probability it is within $1 \s$, 
and in 6\% of the cases the error exceeds $3 \s$.

\subsection{The Estimation of the Trigger Efficiency}

We estimated the trigger efficiency assuming that GRBs of near-threshold 
count rate do not differ in their temporal properties from stronger GRBs 
(which is not exactly true). The procedure consists
of the following steps:

1. For a given peak count rate $P_0$, sample randomly one of the stronger bursts
with a  peak count rate $P> P_0$ of the same class (i.e., simple or complex) 
for which we are going to estimate the trigger efficiency and
subtract the background thus extracting a pure signal.

2. Sample randomly another strong event of arbitrary class in order to use its
linear background fit.

3. Rescale the signal by factor $P_0/P$ and distribute it between the two 
detectors having the largest 
projected areas for a randomly sampled direction of the burst. 
Distribute the signal  proportionally 
to their projected areas. This step does not take into account the reflection 
from atmosphere which causes more uniform exposure of the detectors 
thus enhancing
the trigger efficiency. The procedure also ignores the dependence of the 
detector response matrix on the projection angle. This dependence also enhance 
the trigger efficiency. Thus, our procedure slightly underestimates the 
trigger efficiency. 
It also neglects the probability of triggering three detectors for weak bursts.

4. Add a new linear background from two arbitrary detectors of the event sampled
 at step 2
and a corresponding Poisson noise (taking into account that some noise is 
already there in the rescaled signal).

5. Try the trigger procedure to the rescaled burst as it is programmed in BATSE
(with the $5.5\sigma$ threshold that was used most of the observational time,
see Meegan et al. 1998).

The fraction of rescaled bursts triggered with this procedure is the trigger 
efficiency. Note that with such a procedure the trigger efficiency 
is a function of the {\it expected} count rate.

\subsection{Robustness of the ATP: Test for Brightness Dependent Biases} 

Brightness dependent errors in the ATP induced by Poisson noise are:

-- Errors in the peak count rate estimate which is used as the 
normalization of a time profile, 

-- Errors in the peak position, 

-- Trigger selection effects which removes short or short spike dominated 
events from the sample.

\begin{figure}
\centerline{\epsfxsize=11cm {\epsfbox{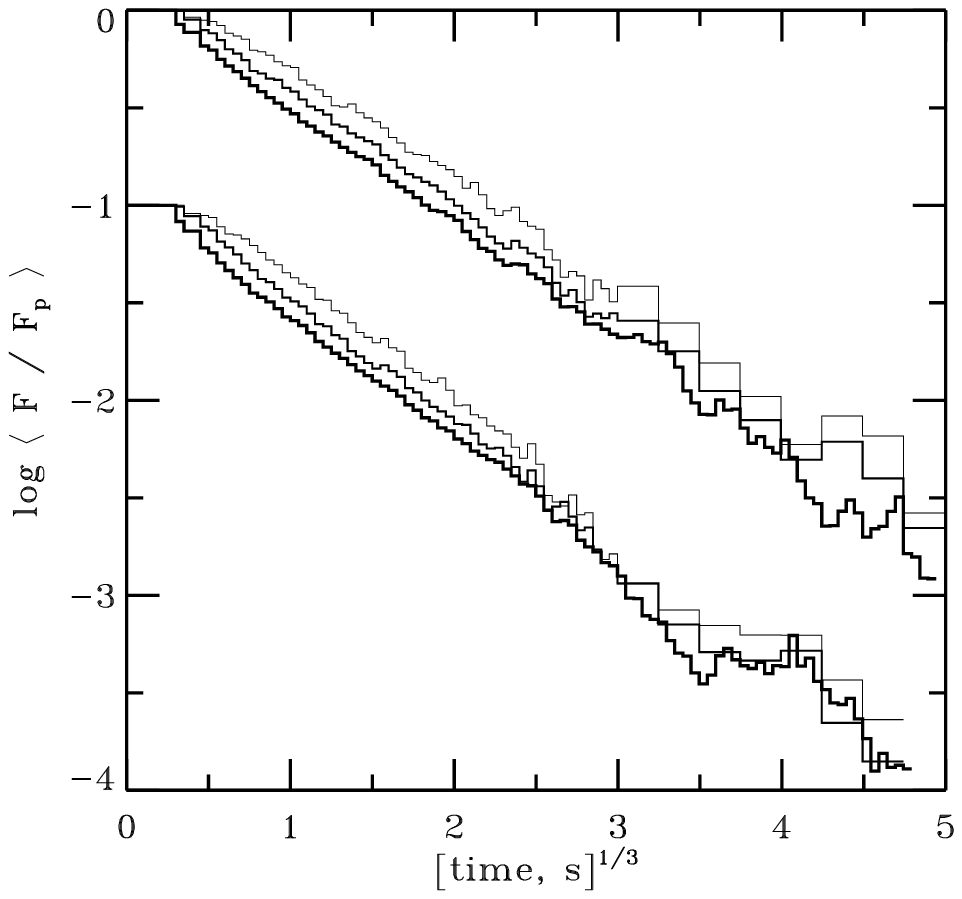}}}
\caption{
Deformations of the decaying (upper histograms) and 
rising (lower histograms) slopes of the ATP after rescaling of the parent 
sample of bright bursts. 
Thickest histograms: the ATP of the parent sample.
Medium thick histograms: the ATP of the parent sample rescaled to 
a peak count rate of 150 counts/64 ms.
Thin histograms: the ATP of the parent sample rescaled to 
a peak count rate of 75 counts/64 ms. 
The ATP of the parent sample rescaled to  
250 counts/64 ms is almost indistinguishable from the parent sample. 
\label{fig9}} 
\end{figure}

We estimated the brightness dependent deformations of the ATP by rescaling 
strong 
events to low peak count rates and applying simulated trigger selection to the 
rescaled sample. The parent sample included 353 GRBs with the highest peak 
count rates of all morphologies and durations. The results are presented in 
Table~\ref{tab4} and in Figure~\ref{fig9}. 


\begin{table}[htbp]
\caption{Time constants measured with $\nu=1/3$ for different peak count rate} 
\begin{center}
\begin{tabular}{l c c c}
\tableline
Peak counts  &    $t_r$ &   $t_d$ &   Trigger efficiency \\ 
\tableline
Parent sample &    0.32 &     0.43      & 1.00 \\ 
250           &    0.31 &     0.42      & 0.97\\ 
150           &    0.31 &     0.42      & 0.83\\ 
110           &    0.32 &     0.44      & 0.71\\ 
75            &    0.33 &     0.47      & 0.57\\ 
\tableline
\end{tabular}
\end{center}
\tablecomments{
Test for robustness of ATP slopes against brightness dependent effects.
The parent sample consists of 353 GRBs with the peak count rate (in 64ms resolution) 
in the interval $500 - \infty$. 
Then it was rescaled to the count rates displayed in the first column. Note that
all time constants are correlated as all of them have the same parent sample.
Therefore the errors of Table 1 are not applicable here and the time constants 
for different peak count rates have small dispersion. When comparing 
with Table 1 one can use the approximate coefficient 0.0072 to translate
peak count rates into peak fluxes with units $\ph\secinv\cminvsq$. 
}
\label{tab4} 
\end{table}

We can state that down to the sample 5 of Table 1 deformations of the ATP are 
negligible (the count rate 110 counts/64 ms is 
near the boundary between samples 5 and 6)
and only for sample 6 (75 counts/64 ms) are they significant. Therefore we 
can measure the variation of the ATP slopes without the 
rescaling procedure which would have erased some information for the strong 
samples.

\section{Stretched Exponential Fits and Their Errors}

 As was noted in Stern (1996), the main problem when fitting ATPs 
is the correlation of deviations in different time bins 
(the ATP is the sum of more or less smooth but very different curves). 
This means that 
one cannot rely on deviations of individual profiles in a $\chi^2$ fit, 
especially when estimating the accuracy of this fit. In principle, a proper 
solution of the problem should exist and it should require an overall 
correlation matrix implemented into the maximum likelihood method, but we
expect that such a solution is not an easy one. In Stern (1996), 
the errors were estimated using a number of smaller samples of bursts. 
The variance of the time constants derived from the smaller samples was 
then rescaled to larger samples as 
$1/\sqrt{N}$, where $N$ is the number of bursts in a sample.
With a limited statistics  such a procedure can only give a very approximate 
estimate of the errors and is
very unreliable if one  uses it to estimate the statistical 
significance level of observed effects.

Following SPS97 we estimate the statistical errors using the pulse 
avalanche model simulations. Here we present further details of the procedure. 
For a description
of the pulse avalanche model, see Stern \& Svensson (1996).
The general time-scale in the model is defined mainly by  
an upper cutoff for the pulse width distribution. It was tuned to get desirable 
time constants for the stretched exponential average time profiles.
Another parameter that was varied is the ``criticality index'', $\mu$, 
 which defines the Poisson average of baby pulses per parent pulse in 
the avalanche. At supercritical values of $\mu$, the process diverges. 
Varying this index some finer features like the stretched exponential index and
the rise/decay asymmetry of the average time profile could be tuned. 
We, however, varied $\mu$ mainly in order 
to test a possible model-dependency of the statistical errors.

We found that the rise/decay asymmetry of the ATP can be 
described better when we introduced a ``global envelope'', to be more exact, an 
external time dependence
for the criticality index, $\mu = \mu^0 \exp(-t/T)$, where the ``global'' time
constant $T$ is large ($T$ = 400 -- 600 s). Then the main peak tends to 
appear earlier, the rise/decay asymmetry is therefore enhanced. 
Such a global envelope with
a large time constant seems natural in many scenarios of GRB emission.

To fit the ATP one needs to split the ATP into a number of bins. This number 
should not be too large, otherwise neighboring bins will be too strongly
correlated and the value of $\chi^2$ or of other likelihood estimators would be
completely meaningless. We choose equidistant binning in the $t^{1/3}$ scale,
each bin being 0.25  $\s^{1/3}$ wide. Correlations are still strong, but with a
wider binning we could loose some information. Our value of $\chi^2$  is 
still not usable as a direct estimator of the quality of the fit, but using the 
model we can calculate the distribution of $\chi^2$ for many ``intrinsically
good'' (i.e., giving a good stretched exponential at high statistic)
ATPs and then define the actual effective number of degrees
of freedom and the renormalization coefficient for $\chi^2$. 

When fitting an ATP for a sample of $N$ bursts, we must know how
the deviations in each time bin are distributed for many independent 
samples. 
We calculated such distributions using $N \times K$ simulation runs 
which produce $K$ independent samples of $N$ events each 
(typically $K=500$ and $N$ vary from 100 to 1000). 
We found that the deviations are excellently
described by a ``continuous Poisson'' distribution, or, in other terms, 
by a gamma-distribution: $\Phi(x,a) = a^{x} e^{-a}/\Gamma(x+1)$,  
where $x$ is the distributed
value, and $a$ is a parameter equivalent to the Poisson average (traditionally
gamma-distribution is used with $a$ being the distributed value and $x - a$ a
parameter). 

We found that the Poisson average $a$ in bin $j$ for all studied 
cases can be 
parameterized as $a_j = p_j\; N\; \xi(p_j)\; \phi(p_j N)$ where $p_j$ is the ATP 
value in the bin (averaged over $N \times K$ events), 
$\xi$ and $\phi$ are slowly varying functions: $\xi$ varies between 3 and 5, 
$\phi(x) = 1$ at $x > 2$ and smoothly decreases at $x < 2$.
 Then the standard maximum likelihood procedure was applied with an estimator
$\chi^2 = - 2 \sum_j \ln [\Phi(p_j-h_j,a_j)/\Phi(a_j,a_j)]$, where $h_j$ is our 
hypothesis: $h_j=\beta \exp[-(t/t_0)^{1/3}]$. At large values of $a_j$,  
this expression coincides with the traditional $\chi^2$ estimator.

\begin{figure}
\centerline{\epsfxsize=11cm {\epsfbox{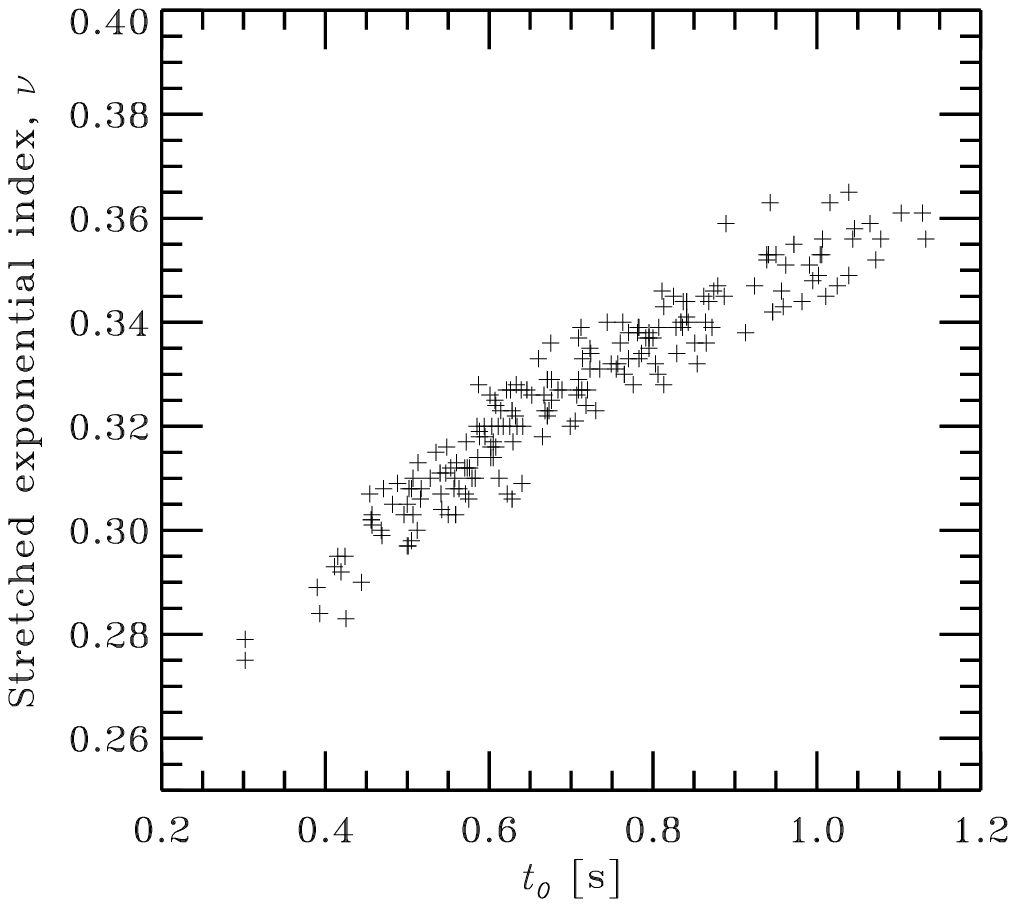}}}
\caption{
Scatter diagram for best fit values of stretched exponential index, $\nu$, and time 
constant, $\t0$, for pulse avalanche simulations. 
\label{fig10}} 
\end{figure}

We need two kind of fits depending on the aim. If we are interested in the 
shape of the ATP slope, we use a 3-parameter fit: 
$f(t)= \beta \exp[-|t/\t0|^{\nu}]$, where $\beta$, $\nu$, and $\t0$ are free 
parameters with the additional requirement that $\beta$ is close to 1 which 
is fulfilled in all reasonable cases.
With pulse avalanche simulation runs (500 $\times$ 300) and (500 
$\times$ 1000) we 
found that the error in the best fit estimate for $\nu$ is: 
$\sigma_{\nu}/\nu$ =0.049 $\times \sqrt{1000/N}$ 
where $N$ is the number of events in the sample.
The cross-correlation of $\t0$ and $\nu$ for a pulse avalanche simulation run
(200 $\times$ 750) is shown in Figure~\ref{fig10}. 
One can see that it is very strong 
and the error for $\t0$ is large for a free $\nu$.

However, if we are interested in the behavior of time constants as a function 
of brightness we should fix $\nu$. Indeed, we can compare time constants 
only with a hypothesis that the ATPs have the same shape (i.e., the same
SE index $\nu$) intrinsically.
Our studies demonstrate that this hypothesis is not exactly true (see \S~4).
Nevertheless, the measurement of $\t0$ with constant $\nu$ is still the best 
that can be done. The deformation of the slope, if moderate, gives a 
second-order error. Therefore we measure $\t0$ with a 2-parameter fit.

In principle, one can set $\beta=1$ and make 
a one parameter fit, but the first bin includes all possible
biases from the finite resolution and the Poisson noise. 
We therefore excluded the first two
bins ($t^{1/3} < 0.5$) from the fit treating $\beta$ as a free parameter.
The upper limit for the fitting range was set to $t^{1/3}=5$ (i.e.,  
$t=125$ s).

As far as we have two slopes of the profile --- pre-peak
and post-peak, we fitted them simultaneously with different time constant, 
$\tr$ (pre-peak, rising slope), $\td$ (post-peak, decaying slope),  
and a common $\beta$. 
Performing a number of model runs with different parameters, we found the
accuracy of the stretched exponential fits to be  almost model-independent. 
The standard
deviation for the time constant does not change more than by 5\% for different 
parameters and scales as $1/\sqrt{N}$ depending on the number of events in the 
sample.
The accuracy slowly increases, when the fitting time interval is extended
(see Table~\ref{tab5}). 
We chose the widest 
interval, which is the default for the results presented elsewhere in the paper.


\begin{table}[htpb]
\caption{Accuracy of the determination of the time constants with 
fixed SE index $\nu$} 
\begin{center}
\begin{tabular}{lc}
\tableline
Fitting interval &  $\sigma(\trd)/(\trd\sqrt{100/N})$ \\ 
\tableline
0.125$<|t|<$8  & 0.252 \\ 
0.125$<|t|<$27 & 0.219 \\ 
0.125$<|t|<$64 & 0.205 \\
0.125$<|t|<$125& 0.201 \\ 
\tableline
\end{tabular}
\end{center}
\label{tab5}
\end{table}

For the sum and ratio of time constants for pre-peak and post-peak slopes
we have:
\bez  
\sigma(\tr+\td)/(\tr+\td) &=&0.196 \sqrt{100/N}, \\ 
\sigma(\td/\tr)/(\td/\tr) &=&0.135 \sqrt{100/N}.  
\eez  
Note, that the relative accuracy for the sum of time constants is close to that
for one  constant while the accuracy for their ratio is considerably better.
This results from the strong correlation between the two slopes -- 
a circumstance that
favors the measurement of shape - brightness correlations and complicates 
the measurement of a time dilation effect.

\begin{figure}
\centerline{\epsfxsize=11cm {\epsfbox{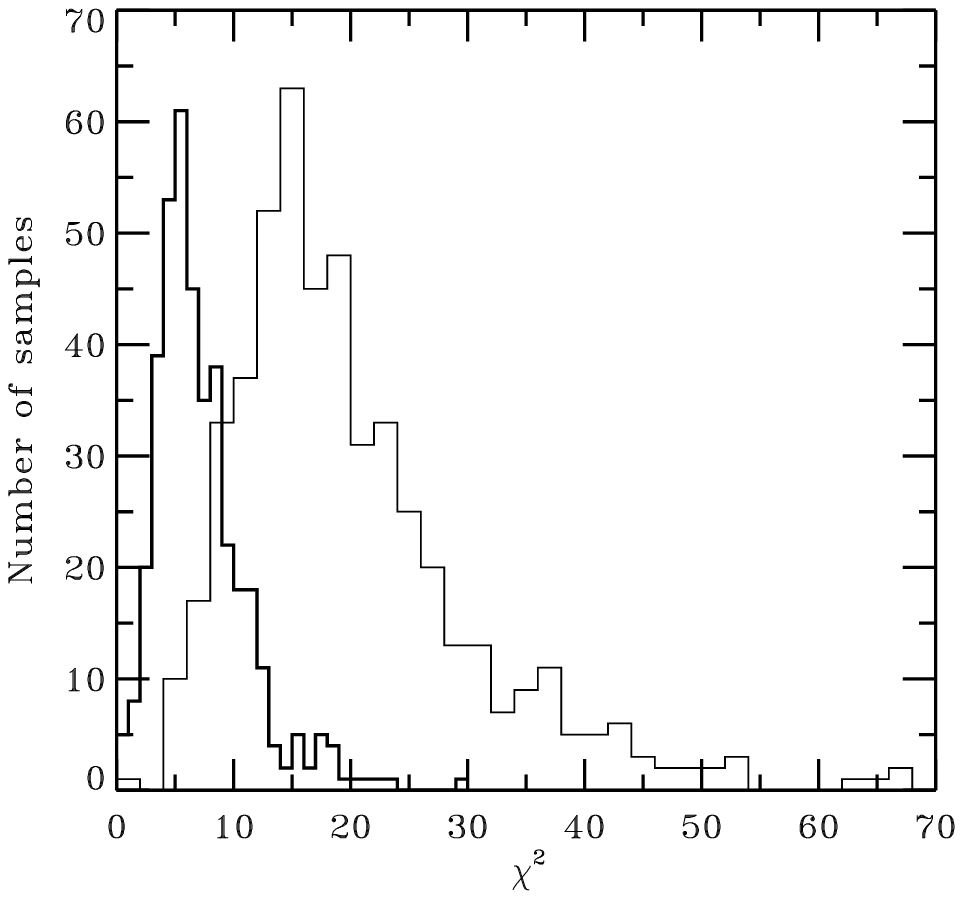}}}
\caption{
Examples of $\chi^2$ distributions for simulated samples of GRBs.
Thick histogram: 
3-parameter fit of the decaying slope (free $\beta, \nu$, and $\t0$),
1000 events in the sample, 15 formal degrees of freedom; thin histogram: 
3-parameter simultaneous fit of both slopes of the ATP (free $\tr, \td$, and 
a common $\beta$; $\nu$ is fixed to 1/3), 300 events in the sample, 
33 formal degrees of freedom.   
\label{fig11}} 
\end{figure} 

As was mentioned above, the formal values of the $\chi^2$ have no direct 
interpretation, the Pearson criterion does not work in this case because of 
strong correlations along the ATP. This could be interpreted as the effective
(unknown) number of degrees of freedom being smaller than the numbers of bins. 
Actually the distribution of $\chi^2$ becomes wider for a larger number of 
events in the sample. In Figure~\ref{fig11}
we present two distributions of $\chi^2$ for
simulated samples which could be used for an approximate evaluation of the 
goodness of 
$\chi^2$ obtained in the ATP fittings of real GRB samples.




\begin{thebibliography}{}
\expandafter\ifx\csname natexlab\endcsname\relax\def\natexlab#1{#1}\fi

\bibitem[{{Abraham}(1997)}]{abr97}
Abraham, R. 1997, Nature, 387, 850 

\bibitem[{{Blinnikov}(1984)}]{blin84}
Blinnikov, S.~I., Novikov, I.~D., Perevodchikova, T.~V., Polnarev, A.~G. 
1984, Sov. Astron. Lett., 10, 177 

\bibitem[{{Brainerd}(1994)}]{bra94}
Brainerd, J. J. 1994, \apjlett, 428, L1 

\bibitem[{{Ching}(1991)}]{ching91} 
Ching, E. 1991, Phys. Rev. A, 44, 3622 

\bibitem[{{Fenimore \& Bloom}(1995)}]{febl95}
Fenimore, E.~E., \& Bloom, J.~S., 1995, \apj, 453, 25 

\bibitem[{{Hartwick \& Schade}(1990)}]{har90}
Hartwick, F.~D.~A., \& Schade, D. 1990, ARA\&A, 28, 437 

\bibitem[{{in 't Zand \& Fenimore}(1994)}]{zand94}
in 't Zand, J.~J.~M., \& Fenimore, E.~E., 1994, 
in Proc. 2nd Huntsville Gamma-Ray Burst Workshop  
(New York: AIP), 
eds. G.~J. Fishman, J.~J. Brainerd, \& Hurley K., p. 692  

\bibitem[{{Jensen}(1992)}]{jen92} 
Jensen, M.~H.,  Paladin, G., \& Vulpiani, A.  1992, Phys. Rev. A, 45, 7214 

\bibitem[{{Kommers et al.}(1997)}]{kom97}
Kommers, J.~M., Lewin, W.~H.~G., Kouveliotou C., van Paradijs, J., 
Pendleton, G.~N., Meegan, C.~A., Fishman, G.~J. 1997, \apj, 491, 704 

\bibitem[{{Lee \& Petrosian}(1997)}]{lee97}
Lee, T. T., \& Petrosian, V. 1997, \apj, 474, 37 

\bibitem[{{Lestrade}(1994)}]{les94}
Lestrade, J.~P. 1994, \apj, 429, L5 

\bibitem[{{Li \& Fenimore}(1996)}]{life96}
Li, H., \& Fenimore, E. E. 1996, \apjlett, 469, L115 


\bibitem[{{Lipunov et al.}(1995)}]{lip95}
Lipunov, V.~M., Postnov, K.~A., Prokhorov, M.~E., Panchenko, I.~E., \&
Jorgensen, H. 1995, \apj, 454, 593 

\bibitem[{{Madau et al.}(1996)}]{madau96}
Madau, P., Ferguson, H.~C., Dickinson, M.~E., Giavalisco, M., Steidel, C.~C.,
\& Fruchter, A. 1996, MNRAS, 283, 1388 

\bibitem[{{Meegan et al.}(1996)}]{mee96}
Meegan, C. A., et al. 1996, \apjs, 106, 65

\bibitem[{{Meegan et al.}(1998)}]{mee98}
Meegan, C. A., et al. 1998, Current BATSE Gamma-Ray Burst Catalog, 
http://www.batse.msfc.nasa.gov/data/grb/catalog


\bibitem[{{Mitrofanov et al.}(1996)}]{mit96}
Mitrofanov, I. G., Chernenko, A. M., Pozanenko, A. S., Briggs, M. S.,
Paciesas, W. S., Fishman, G. J., Meegan, C. A.,  \& Sagdeev, R. Z. 1996,
\apj, 459, 570

\bibitem[{{Mitrofanov et al.}(1997)}]{mit97}
Mitrofanov, I.~G., Litvak, M.~L., \& Ushakov D.~A. 
1997, \apj, 490, 509 

\bibitem[{{Norris et al.}(1994)}]{nor94}
Norris, J. P., Nemiroff, R. J., Scargle, J. D., Kouveliotou, C., Fishman, G. J.,
Meegan, C. A., Paciesas, W. S., \& Bonnell, J. T.
1994,  \apj, 424, 540

\bibitem[{{Norris et al.}(1996)}]{nor96}
Norris, J.~P., Nemiroff, R.~J.,  Bonnell, J.~T., Scargle, J.~D., 
Kouveliotou, C., Paciesas, W.~S.,  Meegan, C.~A., \& Fishman, G.~J.
1996, \apj, 459, 393

\bibitem[{{Paczy\'nski}(1992)}]{pac92}
Paczy\'nski, B. 1992, Nature, 355, 521 

\bibitem[{{Pendleton et al.}(1997)}]{pen97}
Pendleton, G.~N., et al. 1997, ApJ, 489, 175 

\bibitem[{{Petrosian \& Lee}(1996)}]{pet96}
Petrosian, V., \& Lee, T.~T. 1996, ApJ, 467, L29 

\bibitem[{{Prokhorov et al.}(1997)}]{pos97}
Prokhorov, M.~E., Lipunov, V.~M., \& Postnov, K.~A. 
1997, in Proc. XXXII Rencontres de Moriond, Les Arcs, France, in press
(astro-ph/9704039) 

\bibitem[{{Stern}(1996)}]{ste96}
Stern, B.~E. 1996, \apjlett, 464, L111

\bibitem[{{Stern \& Svensson}(1996)}]{ssv96}
Stern, B.~E., \& Svensson, R. 1996, \apjlett, 469, L109 

\bibitem[{{Stern et al.}(1997)}]{sps97}
Stern, B.~E., Poutanen, J., \& Svensson, R. 1997, \apjlett, 489, L41 
(SPS97) 

\bibitem[{{Stern et al.}(1997b)}]{ssp97}
Stern, B.~E., Svensson, R., \& Poutanen, J. 1997,
in The 2nd INTEGRAL Workshop: The Transparent Universe, St. Malo, France,
Sept 1996, 
ESA SP-382, 473

\end{thebibliography}
\end{document}